\documentclass[3p,twocolumn]{elsarticle}

\usepackage{graphicx}
\usepackage[T1]{fontenc}

\biboptions{sort&compress}

\usepackage[T1]{fontenc}
\usepackage[utf8]{inputenc}

\usepackage[version=3]{mhchem}

\usepackage{graphicx} 

\usepackage[implicit=false,colorlinks=true,urlcolor=blue,bookmarks=false,pdfpagelabels=false]{hyperref}

\begin{document}

\title{
The DNA Nucleobase Thymine in Motion -- Intersystem Crossing Simulated with Surface Hopping
\tnoteref{t1}
} 
\tnotetext[t1]{This paper is dedicated to Professor Lorenz S.\ Cederbaum on the occasion of his 70th birthday.}

\author[UoV]{Sebastian Mai}
\author[UoV]{Martin Richter\fnref{fn1}}
\author[UoV]{Philipp Marquetand\corref{cor1}}
\ead{philipp.marquetand@univie.ac.at}
\author[UoV]{Leticia Gonz\'alez\corref{cor1}}
\ead{leticia.gonzalez@univie.ac.at}

\cortext[cor1]{Corresponding author}
\fntext[fn1]{Present address: Max-Born-Institute for Nonlinear Optics and Short Pulse Spectroscopy, 12489 Berlin, Germany.}

\address[UoV]{Institute of Theoretical Chemistry, Faculty of Chemistry, University of Vienna, W\"ahringer Str. 17, 1090 Vienna, Austria.}

\date{\today}

\begin{abstract}
We report ab initio excited-state dynamics simulations on isolated thymine to investigate the mechanism of intersystem crossing, based on CASSCF potential energy surfaces and the \textsc{Sharc} surface hopping method.
We show that even though $S_2 \rightarrow S_1$ internal conversion is not described accurately with CASSCF, intersystem crossing can be correctly simulated. 
Intersystem crossing in thymine occurs from the $S_1$ ($^1n\pi^*$) minimum, via a nearby crossing with $T_2$ ($^3\pi\pi^*$). 
The system further relaxes via ultrafast internal conversion in the triplet manifold to the $T_1$ ($^3\pi\pi^*$) state.
The simulations reveal that, once the system is trapped in the $^1n\pi^*$ minimum, intersystem crossing might proceed with a time constant of 1~ps. 
Furthermore, the change of the system's electronic state is accompanied respectively by elongation/shortening of specific bonds, which could thus be used as indicators to identify which state is populated in the dynamics.
\end{abstract}


\maketitle


\section{Introduction}

DNA and RNA are among the primary absorbers of UV light in all known organisms, stemming from the large absorption cross-section of the nucleobases which are part of DNA and RNA strands.
Absorption of UV light by the nucleobases leads to the formation of excited electronic states, which for this class of compounds mostly deactivate to the electronic ground state within a few ps \cite{Crespo-Hernandez2004CR,Middleton2009ARPC,Kleinermanns2013IRPC,Barbatti2015_1,Barbatti2015_2}.
A very small fraction of excitations of nucleobases in DNA leads to the formation of photochemical products called photolesions, which constitute damage to DNA/RNA and interferes with normal cellular processes.
Among the photolesions, the dimerization of thymine to generate cyclobutane pyrimidine dimers is the most common one \cite{Mouret2006PNAS, Beukers1960BBA, Setlow1962PNAS, Setlow1963S}.
Because of this relevance, the photophysical and photochemical properties of thymine were studied intensively in the last decades.

Among the most controversial aspects of thymine's excited-state dynamics is the importance of intersystem crossing (ISC).
ISC is the radiationless transition between states of different multiplicities, in particular from the initially populated singlet states to triplet states.
Due to the high reactivity and long lifetime of triplet states, once formed, those states could be involved in the formation of photolesions like cyclobutane pyrimidine dimers and pyrimidine 6-4 pyrimidone adducts, as was suggested in the literature \cite{Kwok2008JACS}.
However, other authors have argued that these photolesions are formed without the involvement of triplet states \cite{Schreier2007S,Schreier2009JACS}, or at least that triplet states only marginally contribute to these reactions \cite{Banyasz2012JACS, Liu2016JPCB}.
For these reasons, it is interesting to study the photophysics of thymine.
Besides studies in biological environments, also photophysical investigations in solution and in the gas phase are important here, since they allow separating the intrinsic dynamics of thymine from the effect of the surrounding.

Measurements in aqueous solutions find that ISC yields are 0.004 \cite{Johns1971JACS} to 0.006 \cite{Salet1975PP}, while in chloroform the yield is 0.08 \cite{Rottger2016FD}.
In acetonitrile the reported values range from 0.06 \cite{Salet1975PP} to 0.18 \cite{Lamola1966S}, suggesting that ISC in thymine is solvent dependent and that less polar solvents enhance ISC.
More recently, much effort has been devoted to use time-resolved experimental methods to probe the ultrafast dynamics of thymine.
Gas phase pump-probe experiments are reported by several groups \cite{Kang2002JACS, Ullrich2004PCCP, Canuel2005JCP, He2003JPCA, He2004JPCA, Samoylova2008CP, Kunitski2011CPC,McFarland2014NC}; mostly a sub-ps and a few-ps (5-7 ps) time constant were found, some groups also reported a ns time constant \cite{Kang2002JACS,He2003JPCA,He2004JPCA,Samoylova2008CP,Kunitski2011CPC}.
In aqueous solution, Hare et al.\ have shown that thymine decays biexponentially with two time constants of 2.8 and 30 ps \cite{Hare2007PNAS}.

Nevertheless, an assignment of the experimental time constants to either internal conversion or ISC is difficult solely based on the experimental data and hence the last years have seen a large number of theoretical calculations on thymine.
A number of authors have optimized excited-state minima and crossing points, as well as calculated various paths between these geometries \cite{Perun2006JPCA, Zechmann2008JPCA, Hudock2007JPCA, Gonzalez-Vazquez2009JCCP, Szymczak2009JPCA, Etinski2009JPCA, Serrano-Perez2007JPCB,Asturiol2009JPCA, Merchan2006JPCB, Yamazaki2012JPCA}.
Some authors have also investigated possible stationary pathways for intersystem crossing \cite{Gonzalez-Luque2010JCTC, Serrano-Perez2007JPCB}, showing that ISC is feasible through several singlet-triplet crossing points along its relaxation pathway.
Furthermore, a number of non-adiabatic dynamics studies were performed \cite{Hudock2007JPCA, Szymczak2009JPCA, Lan2009JPCB, Nakayama2013JCP, Asturiol2009JPCA, Asturiol2010PCCP, Barbatti2010PNAS}, but none included the possibility of ISC.

Thus, a logical next step in the investigation of intersystem crossing in thymine is to perform non-adiabatic dynamics simulations including singlet and triplet states with the possibility of ISC.
The \textsc{Sharc} (Surface Hopping including ARbitrary Couplings) excited-state dynamics methodology \cite{Richter2011JCTC,Mai2015IJQC,Mai2014SHARC} is especially well suited for this application, and has already been used successfully for describing ISC in other pyrimidine nucleobases \cite{Mai2013C, Richter2014PCCP, Mai2016JPCL}.


\section{Methodology}

We performed non-adiabatic molecular dynamics simulations on thymine using the \textsc{Sharc} methodology \cite{Richter2011JCTC,Mai2015IJQC,Mai2014SHARC}.
The \textsc{Sharc} method is an extension of Tully's fewest switches surface hopping \cite{Tully1990JCP}, allowing to include in the simulations of all kinds of electronic couplings between the states, in particular spin-orbit couplings which enable ISC.
The electronic Hamiltonian matrix involving the singlet and triplet states of interest (those states are eigenstates of the molecular Coulomb Hamiltonian, hence we denote these states as ``MCH'' states, see also Ref.~\cite{Mai2015IJQC}) including spin-orbit matrix elements is diagonalized to obtain spin-mixed, fully adiabatic states (called ``diagonal states'' in the following).
The surface hopping procedure is then performed on these diagonal states, with the gradients and surface hopping probabilities in the diagonal basis obtained as described in Ref.~\cite{Mai2015IJQC}.
This methodology allows to treat internal conversion and ISC on the same footing, giving a balanced description of all non-radiative processes.

The dynamics simulations were based on SA(4+3)-CASSCF(12,9)/6-31G* (state-averaging over 4 singlets and 3 triplets, complete active space self-consistent field with an active space of 12 electrons in 9 orbitals) calculations performed with \textsc{Molpro} 2012 \cite{Molpro2012}.
The active space contained 8 $\pi/\pi^*$ orbitals and the lone pair of oxygen O$_4$ (ortho to the methyl group), while the lone pair of the other oxygen atom was excluded from the active space, since the $n\pi^*$ states involving excitation from this orbital are very high in energy.

The CASSCF method generally offers a good compromise between accuracy and performance, allowing to simulate an ensemble of sufficient size to sample the ISC channel of the thymine dynamics.
However, it is known that CASSCF does not describe all aspects of the excited-state potential energy surfaces (PES) of thymine correctly \cite{Asturiol2009JPCA,Szymczak2009JPCA,Hudock2007JPCA, Yamazaki2012JPCA,Nakayama2013JCP}.
In particular, CASSCF tends to significantly overestimate the energy of the $^1\pi\pi^*$ state (see Table S1 in the Supplementary Information (SI)), which not only affects the Franck-Condon region, but also the minimum \cite{Yamazaki2012JPCA} and the corresponding conical intersections; consequently, CASSCF will not be able to accurately describe the details of the $^1\pi\pi^*$ deactivation.
Hence, here we will focus on the mechanism which brings about ISC.
An investigation of this mechanism is possible with the CASSCF method, since the vicinity of the $S_1$ ($^1n\pi^*$) minimum and the crossings with the two lowest triplet states $T_1$ and $T_2$ are described properly on CASSCF level of theory, as compared to more accurate CASPT2 (Complete Active Space Perturbation Theory of 2nd order) \cite{Serrano-Perez2007JPCB}, as will be shown below (see also the SI for a comparison of the two levels of theory).

Even though our primary focus is on the ISC mechanism of thymine, we simulated the full dynamics based on excitation to the bright states in the Franck-Condon region, in order to facilitate comparison with previous CASSCF dynamics studies on thymine \cite{Hudock2007JPCA,Szymczak2009JPCA,Barbatti2010PNAS,Asturiol2009JPCA} and uracil \cite{Fingerhut2014JCTC,Richter2014PCCP}.
The initial conditions for the dynamics simulations were sampled from a Wigner distribution around the $S_0$ minimum geometry based on a frequency calculation at the SS-CASSCF(12,9)/6-31G* level of theory.
The initial excited state for each initial condition was determined stochastically, as proposed in Ref.~\cite{Barbatti2007JPPA}.
Based on these initial conditions, an ensemble of 150 trajectories---107 trajectories starting from $S_2$ and 43 from $S_3$---was propagated for up to 2 ps (less if a trajectory relaxed earlier to $S_0$ or $T_1$) with a timestep of 0.25 fs.
Gradients and non-adiabatic coupling vectors were computed for states which are closer than 0.02~a.u.\ to the currently active state.

We note that even though the dynamics simulations were performed in the basis of the diagonal states (4 singlets and 3$\times$3 triplets, giving 13 states in total), final analysis of the populations employed a transformation to the MCH states, which are easier to interpret, since they have defined multiplicities.
Due to the large $S_1-S_2$ energy gap, during the simulations the $S_1$ is $^1n\pi^*$ and $S_2$ is $^1\pi\pi^*$, so these labels can be used interchangeably.
However, the two triplet states of $^3n\pi^*$ and $^3\pi\pi^*$ character often cross in the dynamics, so that there is no one-to-one correspondence to the states $T_1$ (energetically lower triplet state) and $T_2$ (upper triplet state).


\section{Results and Discussion}

\begin{figure}
\centering
 \includegraphics[scale=1]{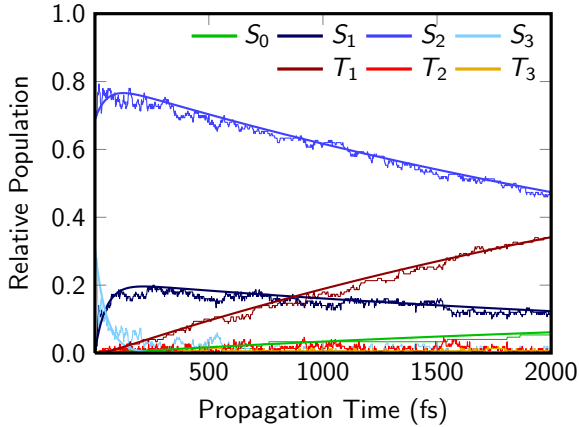}
 \caption{Population of MCH states of thymine averaged over 150~trajectories. The fitted populations based on the kinetic model described in the text are also shown.} 
 \label{fig:pop-t}
\end{figure}

Figure~\ref{fig:pop-t} shows the time-dependent populations of the MCH states averaged over the full ensemble as well as the curves resulting from a global fit, which will be described below.
The figure shows that initially both the $S_2$ ($\pi\pi^*$) and $S_3$ ($\pi\pi^*$) are populated, but the $S_3$ ($\pi\pi^*$) is rapidly depopulated and does not play a significant role in the dynamics after about 200 fs.
Population transfer from $S_2$ ($\pi\pi^*$) to $S_1$ ($n\pi^*$) is much slower and consequently the $S_2$ ($\pi\pi^*$) population reaches about 80\% after a few fs.
Interestingly, the initial population transfer rate from $S_2$ ($\pi\pi^*$) to $S_1$ ($n\pi^*$) is quite large, and within 70 fs 20\% of the total population reaches the $S_1$.
However, for later times the transfer rate becomes much smaller, and even after 2 ps almost 50\% of the total population resides in the $S_2$ ($\pi\pi^*$).
It appears that $S_2$ ($\pi\pi^*$) $\rightarrow$ $S_1$ ($n\pi^*$) population transfer might occur via two channels, where the faster channel is quickly closed as the dynamics proceeds.
From $S_1$ ($n\pi^*$), two relaxation pathways are available: relaxation to the $S_0$ and ISC to the triplet manifold with subsequent relaxation to $T_1$.
Based on the populations, ISC seems to be the more competitive pathway, since within 2 ps the $T_1$ population increases to about 35\%, while the $S_0$ population only reaches 5\% (8 trajectories).
These findings are qualitatively similar to our previous results on uracil, which also shows the mentioned two-step decay of the $S_2$ ($\pi\pi^*$) and more ISC than relaxation to the ground state \cite{Richter2014PCCP}.

\begin{figure}
\centering
 \includegraphics[scale=1]{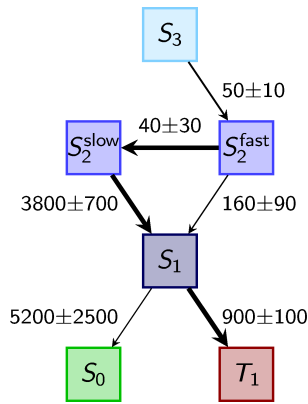}
 \caption{
  Assumed kinetic model of thymine relaxation, fitted time constants, error estimates and resulting extent of population transfer (indicated by arrow thickness). 
  } 
 \label{fig:kinetic}
\end{figure}

In order to obtain rate constants for the population transfer in thymine, a global fit procedure was performed.
The global fit is based on the kinetic model shown in figure~\ref{fig:kinetic}, including 6 species ($S_3$, fast $S_2$, slow $S_2$, $S_1$, $S_0$ and $T_1$) and six time constants ($S_3\rightarrow S_2^\mathrm{fast}$, $S_2^\mathrm{fast}\rightarrow S_2^\mathrm{slow}$, $S_2^\mathrm{fast}\rightarrow S_1$, $S_2^\mathrm{slow}\rightarrow S_1$, $S_1\rightarrow S_0$ and $S_1\rightarrow T_1$); see also the Supplementary Information for more details regarding the fitting procedure.
Since the populations of $T_2$ and $T_3$ are always very small (on average, the sum of their populations is 2\%), they were neglected in the global fit procedure (but note that the $T_2$ is very important for the ISC mechanism, just $T_2\rightarrow T_1$ IC is extremely fast and hence $T_2$ does not acquire a sizeable population).
A simpler model ($S_3\rightarrow S_2\rightarrow S_1\rightarrow S_0$ and $S_1\rightarrow T_1$) was also tested but fails to describe the fast initial rise of the $S_1$ population, so that the more complex model with two paths from $S_2$ to $S_1$ had to be employed.
The fitted time constants are given in Figure~\ref{fig:kinetic}, showing that $S_3\rightarrow S_2$ has a time constant of 50 fs, $S_2\rightarrow S_1$ has two time constants of 160 and 3800 fs, relaxation to $S_0$ has 5200 fs and ISC has 900 fs.
The time constant for $S_2^\mathrm{fast}\rightarrow S_2^\mathrm{slow}$ (40 fs) describes the rate with which the fast $S_2\rightarrow S_1$ channel is quenched.
The error estimates given in Figure~\ref{fig:kinetic} were obtained from bootstrap resampling of the ensemble populations and fitting the kinetic model to each resample \cite{Nangia2004JCP}.

The employed kinetic model also can be beneficially used to fit the data from our previous simulations of uracil~\cite{Richter2014PCCP} (see the SI for the data and fit).
Compared to thymine, in uracil most time constants are slightly faster, with ground state relaxation being much faster at $\tau=$1500~fs.
This difference is most likely due to the missing methyl group of uracil.

Focusing on the results for the singlet multiplicity first, we find that the kinetic model and the fitted time constants agree well with the time constants found in gas phase time-resolved photoelectron experiments, which usually report a fast component around 100 fs and another component of 5-7 ps \cite{Kang2002JACS,Ullrich2004PCCP,Canuel2005JCP,Samoylova2008CP,Kunitski2011CPC,Yu2016PCCP}.
Here, the fast component could be explained with the fast $S_2$ ($\pi\pi^*$) $\rightarrow$ $S_1$ ($n\pi^*$) channel, while the slow component is due to $S_2$ ($\pi\pi^*$) $\rightarrow$ $S_1$ ($n\pi^*$) $\rightarrow$ $S_0$ which involves two steps.
Our simulations agree with other dynamics studies based on CASSCF calculations \cite{Hudock2007JPCA,Szymczak2009JPCA,Barbatti2010PNAS,Asturiol2009JPCA} in the sense that the experimental few-ps time constant is explained by trapping in $S_1$ ($n\pi^*$) and $S_2$ ($\pi\pi^*$).
Different from the other studies, our simulations also predict a fast $S_2$ ($\pi\pi^*$) $\rightarrow$ $S_1$ ($n\pi^*$) relaxation channel, employed by approximately 20\% of the population.

However, it is well known that the CASSCF PES of thymine are qualitatively different from the PES obtained with other levels of theory (e.g., CASPT2 \cite{Serrano-Perez2007JPCB,Yamazaki2012JPCA,Nakayama2013JCP,Buchner2015JACS,Perun2006JPCA,Asturiol2009JPCA,Asturiol2010PCCP}, CC2 \cite{Perun2006JPCA}, DFT \cite{Picconi2011CPC}, or OM2/MRCI \cite{Lan2009JPCB})
Hence, it is questionable whether the observed slow decay from the $S_2$ ($\pi\pi^*$) state to the $S_1$ ($n\pi^*$) state and the sequentiality of $S_2$ ($\pi\pi^*$) $\rightarrow$ $S_1$ ($n\pi^*$) $\rightarrow$ $S_0$ are correct.
Recent MS-CASPT2 studies \cite{Serrano-Perez2007JPCB,Yamazaki2012JPCA,Nakayama2013JCP,Buchner2015JACS, Perun2006JPCA,Asturiol2009JPCA,Asturiol2010PCCP} indicate that a more accurate model for the relaxation of thymine within the singlet manifold involves a portion of the population decaying directly from $S_2$ ($\pi\pi^*$) to $S_0$ (via a CI involving a twist of the C$_5$-C$_6$ bond), while another portion crosses to the $^1n\pi^*$ state and gets trapped.
The finding that there is no trapping in the $^1\pi\pi^*$ state has also been shown experimentally \cite{Hare2007PNAS,McFarland2014NC,Gessner2016ACR}.
The $S_1$ ($n\pi^*$) population then decays on slower timescales to $S_0$, most probably via a recrossing to $S_2$ ($\pi\pi^*$) and relaxation along the same path as the fast decaying portion \cite{Asturiol2009JPCA,Asturiol2010PCCP,Yamazaki2012JPCA}.
The same model was also proposed for uracil \cite{Mercier2008JPCB}.

\begin{figure}[t]
  \centering
  \includegraphics[scale=1]{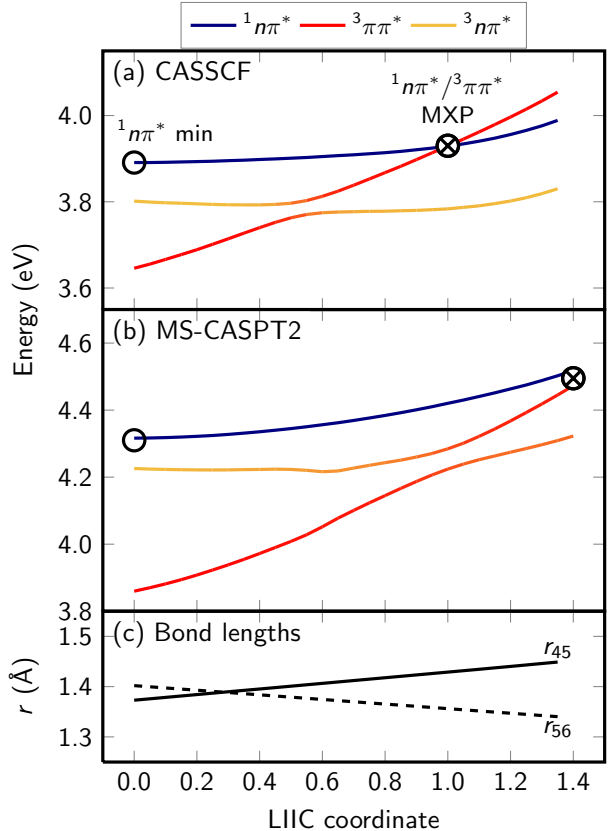}
  \caption{
    LIIC scan from the $S_1$ ($^1n\pi^*$) minimum (empty circle) to the $S_1$/$T_2$ ($^1n\pi^*/^3\pi\pi^*$) minimum energy crossing point (MXP; circle with cross) and linear extrapolation beyond this point. 
    In (a), the energies of $^1n\pi^*$, $^3n\pi^*$ and $^3\pi\pi^*$ along this scan on the dynamics level of theory (CASSCF(12,9)/6-31G*) are shown, in (b) the same states at the MS-CASPT2(12,9)/6-31G* level of theory (at the same geometries). 
    In (c), the corresponding bond lengths of the $C_4-C_5$ and $C_5-C_6$ bonds are plotted.
  } 
 \label{fig:liic}
\end{figure}

There exist some studies where the population trapped in the $S_1$ ($n\pi^*$) minimum has been suggested to be the precursor for ISC \cite{Hare2006JPCB,Hare2007PNAS}.
MS-CASPT2 calculations \cite{Serrano-Perez2007JPCB} have shown that in the vicinity of the $S_1$ ($n\pi^*$) minimum two triplet states are energetically close.
In this region, the $T_1$ state has $^3\pi\pi^*$ character while the $T_2$ state has $^3n\pi^*$ character and thus the $T_2$ PES is almost parallel to the $S_1$ ($n\pi^*$) one (see also Figure~\ref{fig:liic}). 
Close to this region, the characters of the $T_2$ and $T_1$ interconvert due to a $T_2$/$T_1$ ($^3n\pi^*$/$^3\pi\pi^*$) crossing. 
In the vicinity of this coupling region, also a crossing between $T_2$ (here $^3\pi\pi^*$) and $S_1$ ($n\pi^*$) is located.
These two crossings are hence close to the $S_1$ ($^1n\pi^*$) minimum and also to the $^3n\pi^*$ minimum.
The SOC between $S_1$ ($^1n\pi^*$) and $T_2$ ($^3\pi\pi^*$) is reported to be of considerable size (61 cm$^{-1}$) \cite{Serrano-Perez2007JPCB}.
This situation is well reproduced at our CASSCF level of theory, since usually states of $n\pi^*$ character are less affected by the missing dynamical electronic correlation in this method and the $^3\pi\pi^*$ state is fortuitously well-described in the surrounding of the $^{1,3}n\pi^*$ minima.
The SOC in our simulations are also close to the ones reported in the literature \cite{Serrano-Perez2007JPCB}, peaking at 58 cm$^{-1}$ with an average of 34 cm$^{-1}$.

Figure~\ref{fig:liic} shows a linear interpolation in internal coordinates (LIIC) scan from the $S_1$ ($^1n\pi^*$) minimum to the $S_1$/$T_2$ ($^1n\pi^*/^3\pi\pi^*$) minimum energy crossing point on the CASSCF(12,9)/6-31G* level of theory used for the dynamics (starting and end point of the interpolation scan are reported in the SI).
Additionally, a LIIC scan carried out at the MS-CASPT2(12,9)/6-31G* level of theory (using \textsc{Molcas} 8.0 \cite{Aquilante2015JCC} with an IPEA shift of zero, an imaginary level shift of 0.3~a.u., and the same number of states as in the CASSCF calculations) is shown for comparison.
The figure shows that indeed the PES at CASSCF and CASPT2 level agree qualitatively with each other.
Hence, we are confident that CASSCF can describe ISC from the $^1n\pi^*$ minimum in a qualitatively correct way.

The $S_1$/$T_2$ ($^1n\pi^*/^3\pi\pi^*$) crossing shown in Figure~\ref{fig:liic} is responsible for the majority of ISC events in the dynamics simulations. 
We firstly discuss these surface hops in terms of state characters and later on in the frame of MCH states.
Out of 53 trajectories showing ISC, 46 trajectories (87\%) showed a transition of $^1n\pi^*\rightarrow{} ^3\pi\pi^*$ type.
The remaining trajectories showed $^1n\pi^*\rightarrow{} ^3n\pi^*$ transitions (4 times) and ISC from the $^1\pi\pi^*$ state (3 times).
Hence, the results agree with the expectations from the El-Sayed rule \cite{El-Sayed1963JCP}.
The small number of $^1n\pi^*\rightarrow{} ^3n\pi^*$ transitions could be due to the small but non-negligible SOC between these states, or due to mixing of $^3n\pi^*$ and $^3\pi\pi^*$, which would enhance SOC.
When analyzing the ISC transitions in terms of MCH states, we found 13 $S_1\rightarrow T_1$ hops, 37 $S_1\rightarrow T_2$ hops, and 3 $S_{2,3}\rightarrow T_2$ hops.
Thus, in most cases $^1n\pi^*\rightarrow{} ^3\pi\pi^*$ transitions involve as a first step a crossing of $^3\pi\pi^*$ and $^3n\pi^*$ (so that $^3\pi\pi^*$ temporarily becomes $T_2$), followed by $^1n\pi^*\rightarrow{} ^3\pi\pi^*$ ISC, followed by relaxation in the $^3\pi\pi^*$ via a $T_2/T_1$ hop.
The potential energy scans in Figure~\ref{fig:liic} visualize how the $^3\pi\pi^*$ gets destabilized along the LIIC coordinate and first crosses with $^3n\pi^*$ and then with $^1n\pi^*$.
Note that the fast relaxation via the $T_2/T_1$ crossing also explains why the $T_2$ population in Figure~\ref{fig:pop-t} never exceed 5\%.

Thus, in summary it can be said that ISC in thymine occurs close to the $^1n\pi^*$ minimum at a three-state near-degeneracy region, but involves only the $^1n\pi^*$ and $^3\pi\pi^*$ states, whereas the $^3n\pi^*$ state is mostly a spectator.
Other ISC pathways proposed by Serrano-P\'erez et al.\ \cite{Serrano-Perez2007JPCB}, $^1\pi\pi^*\rightarrow{}^3n\pi^*$ and $^1\pi\pi^*\rightarrow{}^3\pi\pi^*$, should play only a minor role in thymine.
For the $^1\pi\pi^*\rightarrow{}^3\pi\pi^*$ pathway, the reported SOC \cite{Serrano-Perez2007JPCB} are only 8 cm$^{-1}$ and the crossing occurs far away from any potential energy minimum, thus allowing ISC only for a limited amount of time.
Consequently, in our simulations we find only a very small number of ISC hops originating from the $^1\pi\pi^*$ ($S_2$) state, as opposed to the larger number of ISC hops from $^1n\pi^*$ ($S_1$).
However, our simulations also overestimate the lifetime of the $^1\pi\pi^*$ ($S_2$) state, and on CASPT2 level the $^1\pi\pi^*$ state is predicted to depopulate very quickly.
Hence, we suggest that ISC in thymine does not originate from the $^1\pi\pi^*$, but from the $^1n\pi^*$ state. 
The latter acts as the doorway state for ISC, as suggested by Hare et al \cite{Hare2006JPCB,Hare2007PNAS}.
We note here that Kunitski et al.\ \cite{Kunitski2011CPC} reported gas phase experiments where they found a 3-exponential decay with time constants of 80$\pm$40 fs, 4.8$\pm$2 ps and 280$\pm$30 ns for thymine.
They concluded that the long lived dark state (lifetime of hundreds of ns) is of triplet character and that the 4.8 ps time constant is related to ISC.
This time constant agrees very well with our simulations---if one assumes a pathway via the ``slow'' $S_2$, the triplet state is populated on a few-ps time scale (3800+900 fs).
Note that according to calculations of Etinski et al.\ \cite{Etinski2009JPCA}, this time scale is sensitive to vibrational excess energy, with a possible range of 9 to 770~ps for ISC from a vibrationally cold $S_1$ state.

Interestingly, in our simulations $S_1$ ($^1n\pi^*$) $\rightarrow$ $T_2$ ($^3\pi\pi^*$) is more than 5 times as fast as $S_1$ ($^1n\pi^*$) $\rightarrow$ $S_0$, which would lead to an ISC yield of about 0.85, much larger than the experimentally measured values \cite{Lamola1966S,Salet1975PP,Johns1971JACS}.
There are several possible reasons for this discrepancy: 
(1) Based on comparison with CASPT2 calculations \cite{Yamazaki2012JPCA}, $S_1$ ($^1n\pi^*$) $\rightarrow$ $S_0$ is most probably too slow on CASSCF level.
(2) ISC might actually be slower than predicted by CASSCF. 
The larger energy difference of the $S_1$ ($^1n\pi^*$) minimum and the $S_1$/$T_2$ ($^1n\pi^*/^3\pi\pi^*$) crossing at CASPT2 (Figure~\ref{fig:liic} (b)) compared to CASSCF (Figure~\ref{fig:liic} (a)) would agree with this thought.
(3) All experimental ISC yields are measured in solution, while our simulations describe the gas phase situation. 
According to many authors, for thymine \cite{Etinski2009JPCA, Gustavsson2006JACS, Serrano-Perez2007JPCB, Yamazaki2012JPCA} and similarly for uracil \cite{Mercier2008JPCB, Kistler2009JPCA, Olsen2010JCTC, DeFusco2011JPCA} a more polar solvent destabilizes the $^1n\pi^*$ state compared to $^1\pi\pi^*$, which could reduce the amount of population going through the $^1n\pi^*$ as well as decrease its lifetime. 
Note that for uracil (essentially thymine without the methyl group), ISC is strongly solvent-dependent, and varies from a few percent in water to about 50\% in aprotic, less polar solvents \cite{Hare2006JPCB,Brister2015JPCL}.
The gas phase can be seen as truely non-polar and an even higher ISC yield could therefore be expected \cite{Mai2016JPCL,Richter2014PCCP}.
Unfortunately, experimental ISC yields for the gas phase are not reported in the literature (neither for uracil nor thymine), according to our knowledge.

\begin{figure*}
  \centering
  \includegraphics[scale=1]{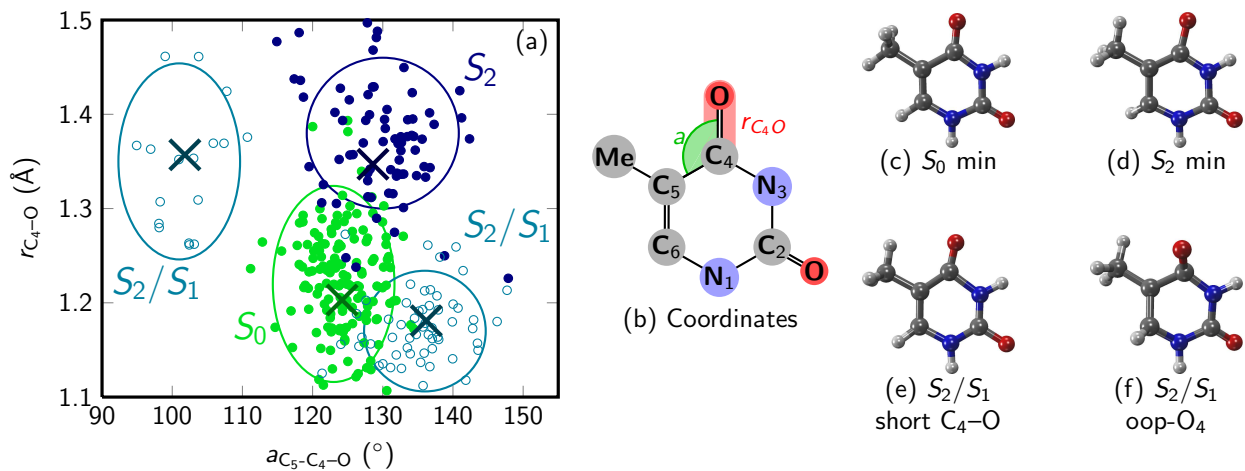}
  \caption{
    Overview over geometric parameters during internal conversion.
    In (a) a scatter plot of the C$_5$-C$_4$--O angle and C$_4$--O bond length for initial geometries ($S_0$, green), for trajectories trapped in $S_2$ ($^1\pi\pi^*$, blue), and for $S_2/S_1$ hopping geometries (open rings).
    The crosses indicate the location of the corresponding minimum or minimum energy crossing point.
    In (b), the two internal coordinates are depicted schematically.
    Panels (c) to (f) show exemplary geometries of $S_0$, $S_2$, and the two different $S_2/S_1$ conical intersections.
  } 
 \label{fig:scatter}
\end{figure*}

Finally, we want to briefly describe the geometric changes seen in the dynamics.
The transition from $^1\pi\pi^*$ ($S_2$) to $^1n\pi^*$ ($S_1$) is mediated by two different types of crossings in our simulations, as shown by the clustering of the hopping geometries in Figure~\ref{fig:scatter}a.
Note that these are the geometries where the trajectories hopped from $S_2$ to $S_1$, and that these are not optimized minimum-energy crossing points. 
However, based on each hopping geometry, an optimization was performed (using the standard algorithm of \textsc{Molpro}\cite{Bearpark1994CPL}), leading to either of the two geometries marked with ``X'' in Figure~\ref{fig:scatter}.
These latter geometries are hence stationary points on the $S_1/S_2$ intersection seam, whereas the hopping geometries are simply geometries close to the seam.
The majority of the $S_2\rightarrow S_1$ hops (54 out of 77) were induced by a crossing whose geometry shows a relatively short C$_4$--O bond length (atom numbering in panel b, optimized energy: 6.60~eV), depicted by the cluster in the lower right corner of Figure~\ref{fig:scatter}a; an exemplary geometry is also shown in panel e.
Since this crossing is relatively close to the Franck-Condon region (middle of Figure~\ref{fig:scatter}a and geometry in c), some trajectories arrive there early in the dynamics (28 trajectories), which is the reason for the fast $S_2\rightarrow S_1$ channel shown in figure~\ref{fig:kinetic}.
The remaining trajectories miss the crossing and get trapped in the $S_2$ minimum with a long C$_4$--O bond (upper right corner, geometry in d).
Trajectories leave the $S_2$ minimum slowly (the slow relaxation channel) and eventually decay to the $S_1$, employing either the ``short C$_4$--O'' crossing or another $S_2/S_1$ crossing.
The latter crossing shows rather long C$_4$--O bonds and small C$_5$-C$_4$--O angles (upper left corner in the figure, optimized energy: 6.30~eV), as well as out-of-plane motion of O$_4$ (``oop-O$_4$''), as can be seen in panel f.
The ``oop-O$_4$'' crossing has been previously reported as ``$^6S_5$'' by Szymczak et al.\ \cite{Szymczak2009JPCA}, whereas the ``short C$_4$--O'' crossing was not reported so far in the literature, to the best of our knowledge.
The two optimized geometries are reported in the SI.
Note, however, that---as mentioned before---the described crossings were obtained at CASSCF level, and that at MS-CASPT2 the $S_2\rightarrow S_1$ relaxation process might be different.

After relaxation to the $^1n\pi^*$ ($S_1$) minimum, a large fraction of the trajectories undergo ISC.
This process is mainly controlled by geometric parameters which strongly destabilize the $^3\pi\pi^*$ state, as was shown in Figure~\ref{fig:liic} for the $C_5-C_6$ and $C_4-C_5$ bond lengths.
A shortening of the $C_5-C_6$ bond for example increases the $^3\pi\pi^*$ energy (see Figure~\ref{fig:liic}), which may facilitate reaching the $^1n\pi^*/^3\pi\pi^*$ crossing.
However, based on the trajectories it is not possible to single out an isolated mode of the molecule which is responsible for $^1n\pi^*\rightarrow{}^3\pi\pi^*$ ISC.
The distribution of geometric parameters of all ISC hopping geometries is not significantly different from the distribution of the same parameters over all trajectories moving in $S_1$.
This suggests that several modes are strongly coupled in bringing about the $^1n\pi^*/^3\pi\pi^*$ crossing and hence ISC.

\begin{figure}
  \centering
  \includegraphics[scale=1]{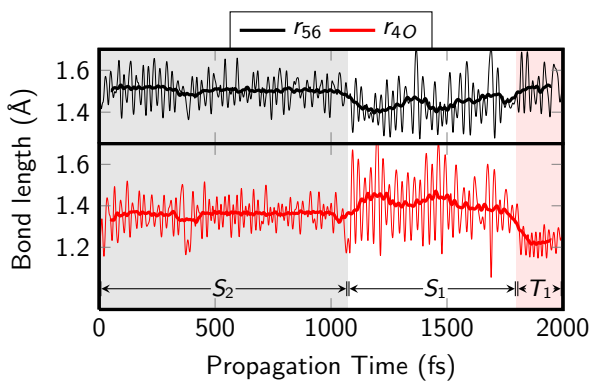}
  \caption{
    Time dependence of the bond lengths of C$_4$--O and C$_5$-C$_6$ of an exemplary trajectory and moving averages (100~fs width). The shaded areas denote time intervals where the trajectory moved in $S_2$, $S_1$ and $T_1$, from left to right. 
  } 
 \label{fig:bonds}
\end{figure}

It was noted for many trajectories that the state hops induce well observable variations to some vibrational modes.
The most prominent of these ``indicator modes'' were the C$_4$--O and C$_5$-C$_6$ bond lengths.
Figure~\ref{fig:bonds} shows for an exemplary trajectory the time dependence of these two bonds, together with the moving averages (an average is taken over 50~fs before and 50~fs after the respective point in time in order to obtain smooth curves, which show the trend on this 100~fs time scale).
The figure also indicates at which times a state-to-state transition in this trajectory occurred.
As can be nicely seen, after excitation to the $S_2$ state the bond length moving averages quickly (within 50~fs) assume the bond lengths of the $S_2$ minimum geometry---1.49~\AA\ for $r_{56}$ and 1.35~\AA\ for $r_{4O}$. 
At 1080~fs, when the trajectory relaxes to the $S_1$ state, the average bond lengths change, with $r_{56}$ becoming shorter (the $S_1$ minimum bond length is 1.42~\AA) while $r_{4O}$ increases (minimum at 1.37~\AA, the larger average in the figure could be explained by a strong excitation of the bond or anharmonicity).
Finally, after ISC to $T_1$ (minimum values: $r_{56}$=1.49~\AA, $r_{4O}$=1.22~\AA) again a significant change of the average bond lengths can be seen, especially for the C$_4$--O bond, which contracted notably.


\section{Conclusion}

Ab-initio non-adiabatic molecular dynamics simulations including singlet and triplet states using the \textsc{Sharc} methodology have been performed to investigate the intersystem crossing mechanism of thymine.
The simulations were based on CASSCF electronic structure calculations.
It was found that after initial photoexcitation to the bright $^1\pi\pi^*$ state the majority of trajectories spends considerable time in this state before relaxing to the $^1n\pi^*$.
From the latter state, the system relaxes slowly (few ps time scale) to the ground state, which agrees with the findings of previous CASSCF-based dynamics studies \cite{Hudock2007JPCA,Szymczak2009JPCA,Barbatti2010PNAS,Asturiol2009JPCA} and yields excited-state lifetimes in agreement with gas-phase pump-probe spectroscopy \cite{Kang2002JACS,Ullrich2004PCCP,Canuel2005JCP,Samoylova2008CP,Kunitski2011CPC,McFarland2014NC,Yu2016PCCP}.
Additionally, it was simulated for the first time that from the $^1n\pi^*$ state efficient ISC on a picosecond time scale occurs, bringing the system to the $^3\pi\pi^*$ state.
Our simulations also show that other ISC pathways described in the literature \cite{Serrano-Perez2007JPCB}, $^1\pi\pi^*\rightarrow{}^3n\pi^*$ and $^1\pi\pi^*\rightarrow{}^3\pi\pi^*$, are not important in the dynamics.

In summary, our results provide evidence that the time-dependent treatment of the relaxation processes in the nucleobase thymine has to include the possibility of ISC in order to fully cover the whole range of important interactions between the different electronic states.
However, it is imperative for a more thorough understanding of thymine's excited-state dynamics that future simulations are based on more accurate electronic structure calculations than CASSCF.
Recent dynamics simulations \cite{Nakayama2013JCP,Buchner2015JACS,Mai2016JPCL} on CASPT2 level of theory are already a step in this direction, but more developments are needed until similar calculations can be carried out routinely.

\section*{Acknowledgments}

Funding from the Austrian Science Fund (FWF), project P25827, and generous allocation of computer ressources at the Vienna Scientific Cluster 2 (VSC2) are gratefully acknowledged. 
We also thank the COST actions CM1204 (XLIC) and CM1305 (ECOSTBio) for support.


\newcommand{\todo}[1]{\textcolor{red}{#1}}
\newcommand{\hereIam}{\noindent\textcolor{black}{\rule{\columnwidth}{7pt}}}

\newcommand{\change}[1]{#1}
\newcommand{\delete}[1]{\ignorespaces}

\newcommand{\changeswitch}{\color{black}}
\newcommand{\changenormal}{\color{black}}


\renewcommand{\textfraction}{0.01}
\renewcommand{\topfraction}{0.9}
\renewcommand{\bottomfraction}{0.8}
\renewcommand{\dbltopfraction}{0.9}
\setcounter{topnumber}{2}
\setcounter{bottomnumber}{2}

\newcommand{\D}{\ensuremath{\mathrm{d}}}

\clearpage

\onecolumn
\begin{center}
{\Large{Supporting Information for: The DNA Nucleobase Thymine in Motion -- Intersystem Crossing Simulated with Surface Hopping}
}
{\small
\\
\vspace{0.1cm}
Sebastian Mai$^a$ , Martin Richter$^{a,1}$ , Philipp Marquetand$^{a,*}$, Leticia Gonz\'alez$^{a,*}$\\
\vspace{0.1cm}
$^{a}$ Institute
of Theoretical Chemistry, Faculty of Chemistry, University of Vienna, Währinger Str. 17, 1090 Vienna, Austria.\\
$^1$ Present address: Max-Born-Institute for Nonlinear Optics and
ort Pulse Spectroscopy, 12489 Berlin, Germany.\\
\vspace{0.1cm}
$^*$ Corresponding author\\
\textit{Email addresses:} philipp.marquetand@univie.ac.at (Philipp Marquetand), leticia.gonzalez@univie.ac.at (Leticia Gonz\'alez)\\
}

\end{center}





\setcitestyle{numbers,open={[S},close={]}}

\setcounter{section}{0}
\setcounter{figure}{0}
\setcounter{table}{0}
\renewcommand{\thefigure}{S\arabic{figure}}
\renewcommand{\thetable}{S\arabic{table}}

\section{Level of Theory Validation}

Table~\ref{tab:validation} presents the energies and energy gaps for several critical points on thymine's potential energy surfaces (PESs).
In the ``CASSCF'' column, the same level of theory as in the dynamics simulations is used.
In the right-most columns, MS-CASPT2 results are presented for comparison.
The latter computations employed \textsc{Molcas} 8.0, a CAS(12,9), the 6-31G* basis set, 4~singlet or 3~triplet states in the state-averaging (SA over singlets and triplets in one calculation is not possible in \textsc{Molcas}), an IPEA shift of zero, and an imaginary level shift of 0.3~a.u.
The same MS-CASPT2 settings were also used for the LIIC scan presented in the main manuscript.
Note that both sets of calculations employed the same geometries, no reoptimization at MS-CASPT2 level was performed.

As can be seen, at the Franck-Condon geometry, CASSCF predicts all vertical excitation energies reasonably well, with the exception of the $S_2$ ($^1\pi\pi^*$) state, which is predicted 2~eV too high.
Also for the minimum geometries, the $S_1$ ($^1n\pi^*$) and $T_1$ ($^3\pi\pi^*$) states are well reproduced, whereas the $S_2$ ($^1\pi\pi^*$) minimum is significantly too high in energy.
Furthermore, at CASPT2 level the $^1\pi\pi^*$ minimum has been located on the $S_1$ surface.\cite{YamazakiSI}

The crossing points in the singlet manifold do not agree well with each other.
However, as the MS-CASPT2 results do not show crossings at these particular geometries, only a proper optimization of the MECPs at MS-CASPT2 level would allow for a full comparison.
Still, the results clearly indicate that some details of the $S_2\rightarrow S_1\rightarrow S_0$ relaxation of thymine might not be well reproduced by our calculations.

Interestingly, the $S_1/T_2$ MECP is reasonably described by CASSCF, as can also be seen in Figure~3 in the main manuscript.
This fact, together with the qualitatively correct description of the $S_1$ ($^1n\pi^*$) minimum, indicates that intersystem crossing should be well described in our simulations.

\begin{table}[!b]
  \centering
  \changeswitch
  \caption{
  \changeswitch
  Energies of critical points at CASSCF and MS-CASPT2 levels of theory.
  Energies are given in eV, relative to the $S_0$ minimum energy (given in Hartree).
  Energy gaps are given in eV.
  For crossings, the average energy of the two involved states is given.
  }
  \label{tab:validation}
  \begin{tabular}{lllll}
    \hline
    Geometry            &\multicolumn{2}{l}{SA-CASSCF(12,9)/6-31G*}     &\multicolumn{2}{l}{MS-CASPT2(12,9)/6-31G*}\\
                        &Energy &Gap &Energy &Gap\\
    \hline
    $S_0$ at $S_0$ min  &-451.7938159     &---  &-452.8009802       &---\\
    $S_1$ ($^1n\pi^*$) at $S_0$ min  &5.14   &---            &5.10   &---\\
    $S_2$ ($^1\pi\pi^*$) at $S_0$ min  &7.20   &---            &5.19   &---\\
    $T_1$ ($^3\pi\pi^*$) at $S_0$ min  &3.99   &---            &3.93   &---\\
    $T_2$ ($^3n\pi^*$) at $S_0$ min  &4.95   &---            &4.96   &---\\
    $S_1$ ($^1n\pi^*$) min           &3.89   &---            &4.18   &---\\
    $S_2$ ($^1\pi\pi^*$) min           &5.85   &---            &4.86   &---\\
    $T_1$ ($^3\pi\pi^*$) min           &3.09   &---            &3.21   &---\\
    $S_2/S_1$ MECP (oop-O$_4$)    &6.30   &$<$0.01        &6.50   &0.24\\
    $S_2/S_1$ MECP (short C$_4$--O)         &6.60   &0.01           &5.47   &0.75\\
    $S_1/S_0$ MECP (boat)         &5.32   &0.03           &5.14   &0.62\\
    $S_1/S_0$ MECP (oop-O$_4$)    &5.46   &$<$0.01        &5.31   &0.34\\
    $S_1/T_2$ MECP                &3.93   &$<$0.01        &4.25   &0.13\\
    \hline
  \end{tabular}
\end{table}

\clearpage

\section{Details on the Kinetic Model Fitting}

The kinetic model used for the population fit is described by the following differential equation system:
\begin{equation}
  \frac{\partial}{\partial t}
  \begin{pmatrix}
    S_3(t)\\
    S_2^\text{fast}(t)\\
    S_2^\text{slow}(t)\\
    S_1(t)\\
    S_0(t)\\
    T_1(t)
  \end{pmatrix}
  =
  \begin{pmatrix}
    -k_{S_3}\\
    +k_{S_3}    &-(k_\text{cool}+k_\text{d.fast})\\
                &+k_\text{cool}           &-k_\text{d.slow}\\
                &+k_\text{d.fast}       &+k_\text{d.slow}       &-(k_\text{Rlx}+k_\text{ISC})\\
                &&                              &+k_\text{Rlx}       &0\\
                &&                              &+k_\text{ISC}       &&0
  \end{pmatrix}
  \cdot
  \begin{pmatrix}
    S_3(t)\\
    S_2^\text{fast}(t)\\
    S_2^\text{slow}(t)\\
    S_1(t)\\
    S_0(t)\\
    T_1(t)
  \end{pmatrix},
  \label{eq:model}
\end{equation}
where $S_3(t)$, $S_2^\text{fast}(t)$, $S_2^\text{slow}(t)$, $S_1(t)$, $S_0(t)$, and $T_1(t)$ are the \emph{model functions} of the respective kinetic species at time $t$.
The $k$'s are the kinetic constants, where $k_{S_3}$ is the decay constant of $S_3(t)$, $k_\text{cool}$ the conversion rate for $S_2^\text{fast}\rightarrow S_2^\text{slow}$, $k_\text{d.fast}$ is the decay constant of $S_2^\text{fast}$ to $S_1$, $k_\text{d.slow}$ is the decay constant of $S_2^\text{slow}$ to $S_1$, $k_\text{Rlx}$ is the $S_1\rightarrow S_0$ constant, and $k_\text{ISC}$ is the $S_1\rightarrow T_1$ constant.

The differential equation system has been solved using the computer algebra system \textsc{Maxima} 5.29.1, yielding the analytical expressions of the six model functions $S_3(t)$, $S_2^\text{fast}(t)$, $S_2^\text{slow}(t)$, $S_1(t)$, $S_0(t)$, and $T_1(t)$.
The initial values are:
\begin{align}
  S_3(0)        &=47/150\nonumber\\
  S_2^\text{fast}(0)&=103/150\nonumber\\
  S_2^\text{slow}(0)&=0\nonumber\\
  S_1(t)&=0\nonumber\\
  S_0(t)&=0\nonumber\\
  T_1(t)&=0,\nonumber
\end{align}
in accordance with the initial distribution of the trajectories.

In a global fitting procedure, using \textsc{Gnuplot} with the Marquardt-Levenberg algorithm, the actual populations from the simulations were fitted to the model functions, as given below:
\begin{align}
  S_3(t)        &\Rightarrow      S_3^\text{simulated}(t)\nonumber\\
  S_2^\text{fast}+S_2^\text{slow}        &\Rightarrow    S_2^\text{simulated}(t)\nonumber\\
  S_1(t)        &\Rightarrow      S_1^\text{simulated}(t)\nonumber\\
  S_0(t)        &\Rightarrow      S_0^\text{simulated}(t)\nonumber\\
  T_1(t)        &\Rightarrow      T_1^\text{simulated}(t),\nonumber
\end{align}
where $A\Rightarrow A^\text{simulated}$ indicates a least-squares fit, where the constants $k_i$ are varied to minimize $\sum_{i} (A(t_i)-A^\text{simulated}(t_i))^2$.

Here, we want to note that according to this procedure, we do not divide the $S_2$ population in the trajectory simulations into subpopulations.
$S_2^\text{fast}$ and $S_2^\text{slow}$ are simply two functions used to model the biexponential decay behaviour of the $S_2$ population.
The actual reason for this behaviour could be that part of the $S_2$ population in our simulations directly moves from the Franch-Condon point to the $S_2/S_1$ crossing, whereas the remaining part of the population moves to the $S_2$ minimum and spends more time until it hits the crossing.

\subsection*{Fitting of the Populations of Uracil}

As mentioned in the main manuscript, the biexponential kinetic model can also be utilized for fitting the excited-state populations
of uracil, from our previous publication~\cite{RichterSI}.
Since we did not report such a fit in that paper, we present the fit here, in order to facilitate comparison with the new thymine results.

\begin{figure}
  \centering
  \includegraphics[scale=1]{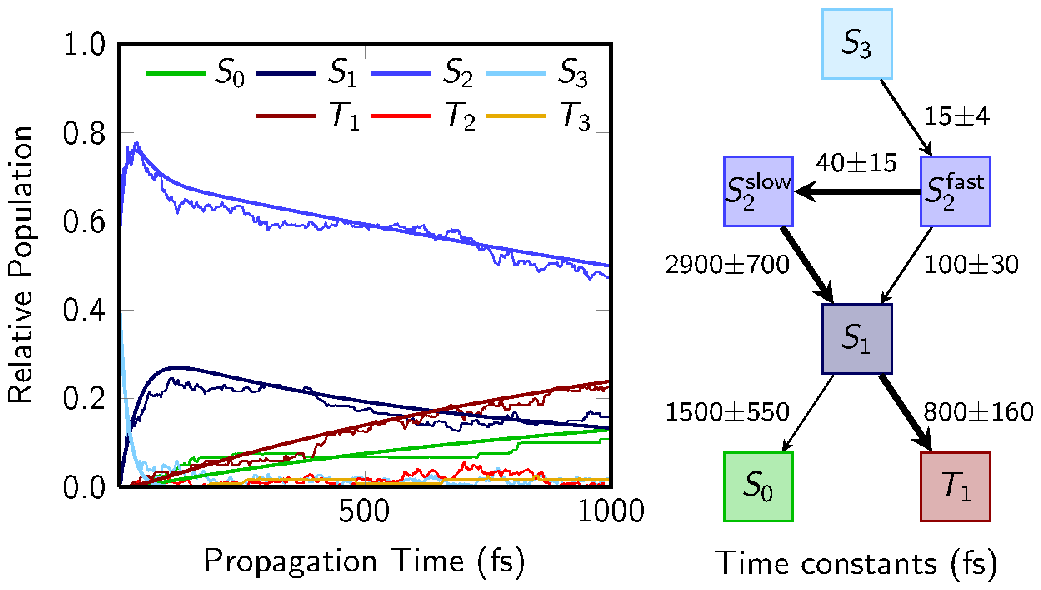}
  \caption{Excited-state populations of uracil (reproduced from Ref.~\cite{RichterSI}, ensemble II with 120~trajectories and CASSCF(14,10)/6-31G*) and fitted time constants with the kinetic model used here for thymine.}
  \label{fig:uracil}
\end{figure}

The populations and time constants are shown in Figure~\ref{fig:uracil}.
For comparison, in our previous publication~\cite{RichterSI} two time constants of 63~fs and 2.8~ps were reported, based on a biexponential fit of the sum of the $S_{\geq 1}$ and $T_{\geq 2}$ states.

\section{Molecular Geometries}

\noindent
\begin{minipage}{0.7\textwidth}
\small
\begin{verbatim}
 15
S1(1npi*) min, SA(4,3)-CASSCF(12,9)/6-31G*
C +0.205075 -0.712113 +0.070989
C -1.063288 -0.207545 +0.190055
C -0.337862 +2.103204 +0.060884
C +1.239085 +0.226715 -0.075745
C +0.474159 -2.197674 +0.049309
N -1.342937 +1.156555 +0.082576
N +0.923657 +1.592693 -0.059017
O -2.196475 -0.971207 +0.201165
O -0.569397 +3.279246 +0.133363
H -2.223892 +1.497048 +0.398503
H +1.645748 +2.273282 -0.113300
H +2.277709 -0.023075 -0.048640
H -0.103455 -2.716011 +0.805912
H +1.521704 -2.395086 +0.243032
H +0.225882 -2.629489 -0.914504
\end{verbatim}
\end{minipage}
\begin{minipage}{0.29\textwidth}
\vspace*{0.5cm}
\includegraphics[width=\textwidth]{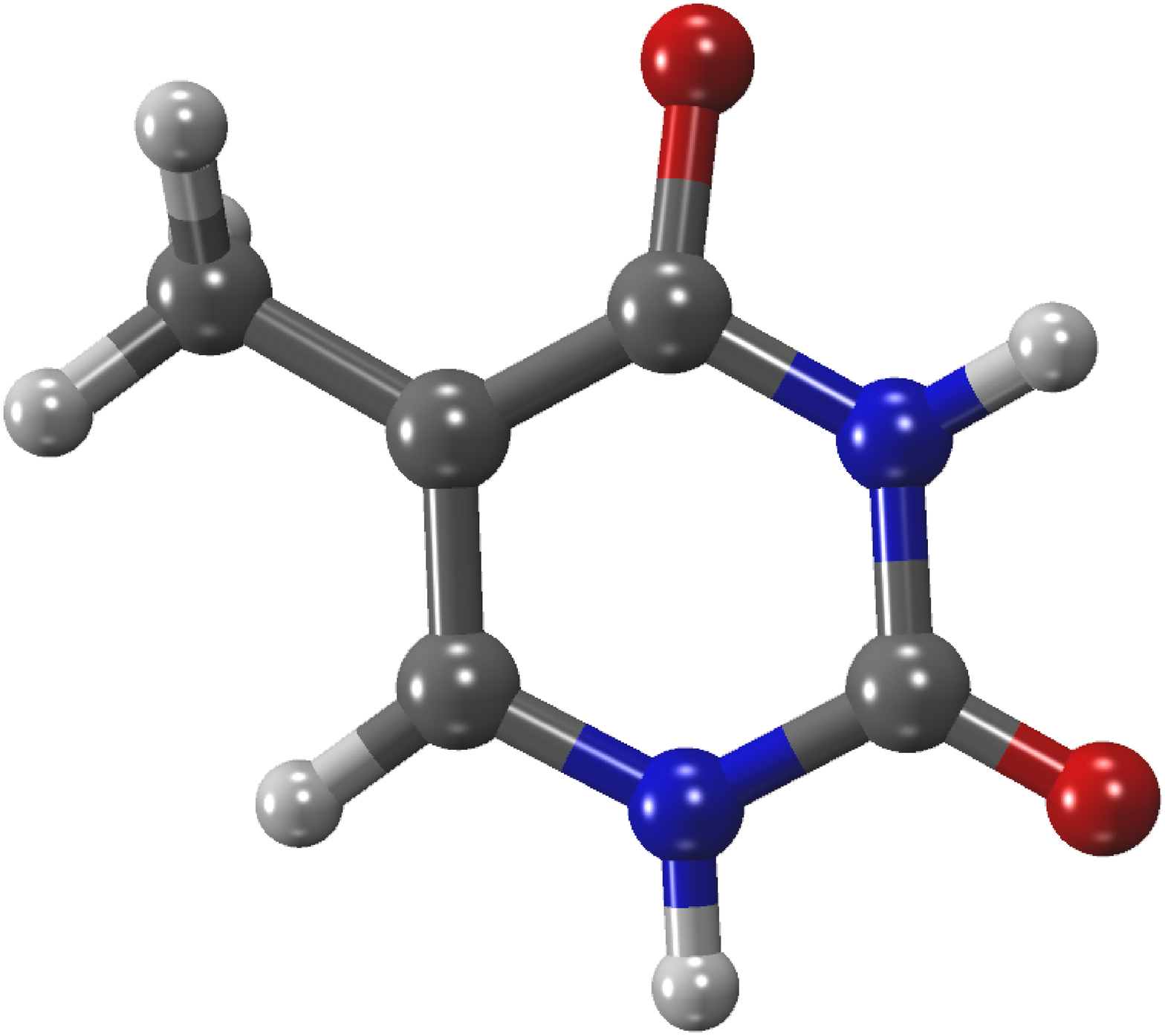}

\includegraphics[width=\textwidth]{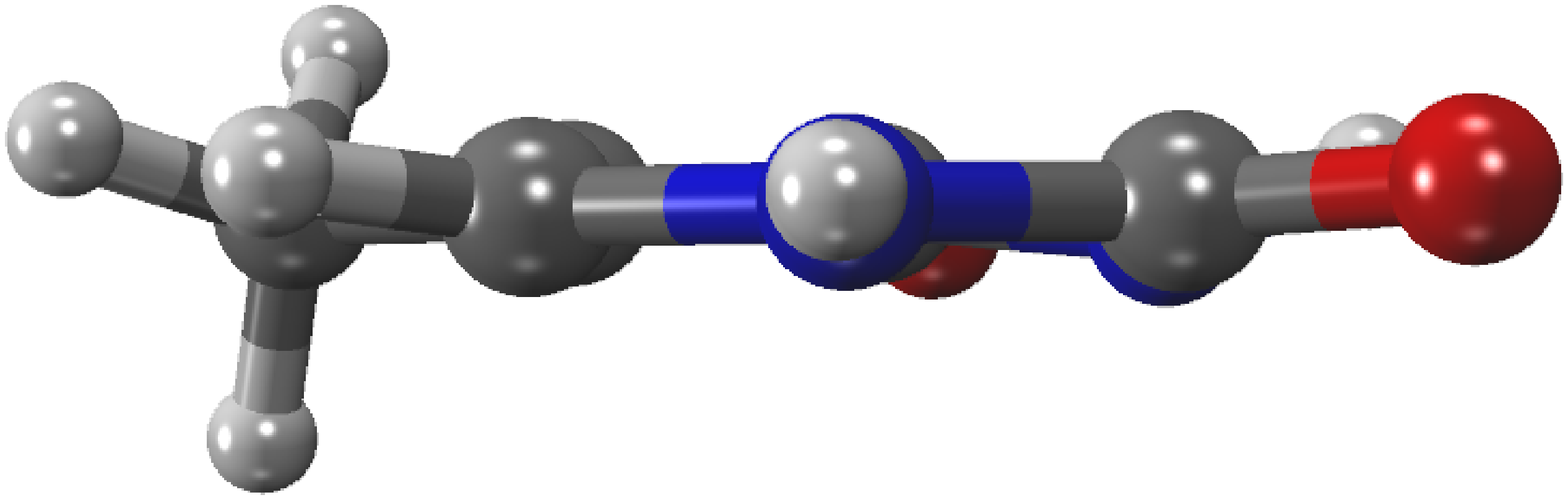}
\end{minipage}

\bigskip
\noindent
\begin{minipage}{0.7\textwidth}
\small
\begin{verbatim}
 15
S1(1npi*)/T2(3pipi*) MXP, SA(4,3)-CASSCF(12,9)/6-31G*
C +0.233164 -0.706194 +0.061724
C -1.086403 -0.206851 +0.269150
C -0.339679 +2.098509 +0.031765
C +1.221543 +0.220027 -0.041484
C +0.481985 -2.192006 +0.012373
N -1.348890 +1.153427 +0.030530
N +0.926517 +1.588056 -0.029929
O -2.173128 -0.970180 -0.094949
O -0.571619 +3.275639 +0.077642
H -2.218321 +1.521499 +0.351851
H +1.655075 +2.262179 -0.069968
H +2.259004 -0.030520 -0.124679
H +0.144294 -2.680543 +0.920435
H +1.540309 -2.404191 -0.105586
H -0.042643 -2.639868 -0.822794
\end{verbatim}
\end{minipage}
\begin{minipage}{0.29\textwidth}
\includegraphics[width=\textwidth]{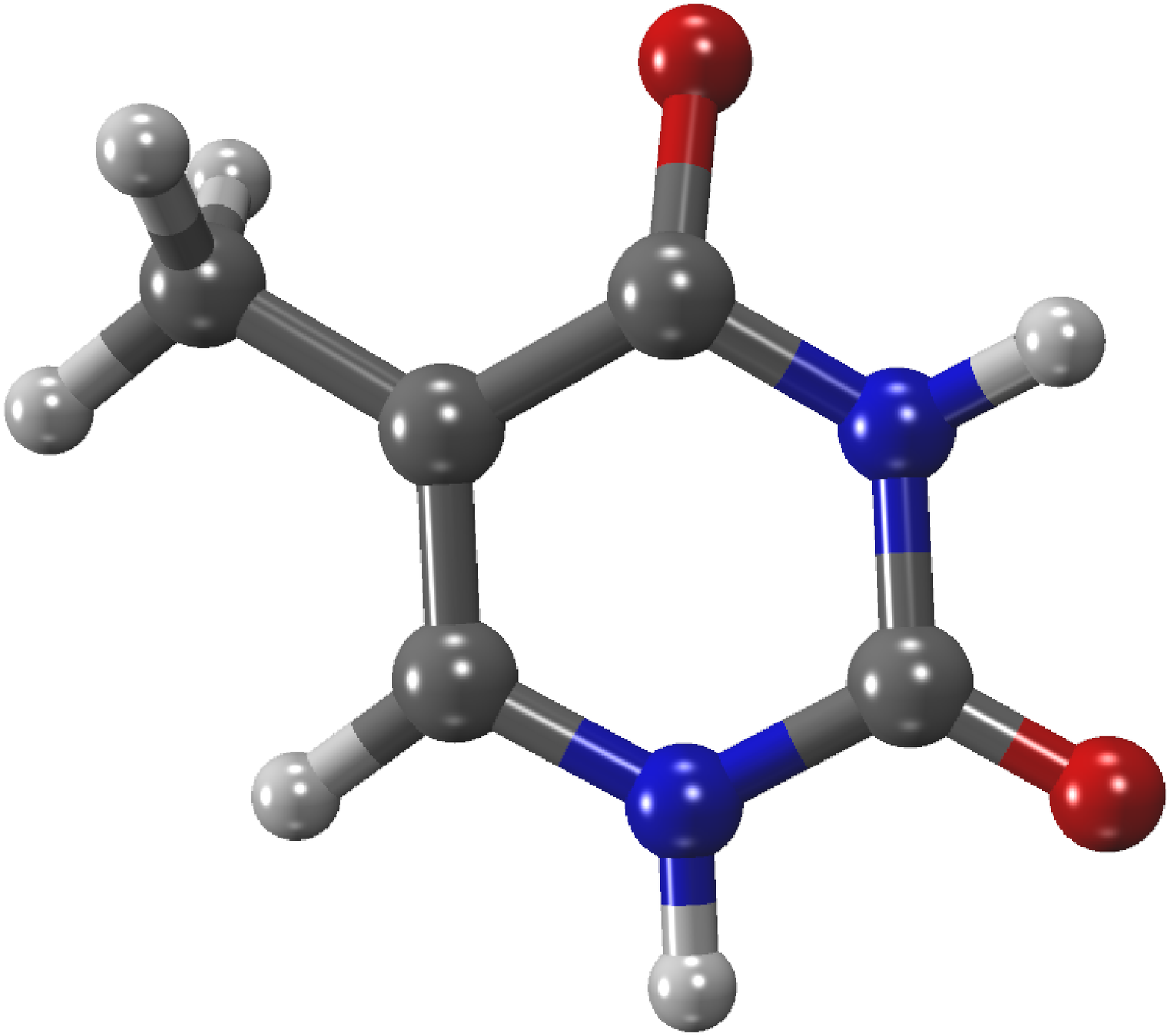}

\includegraphics[width=\textwidth]{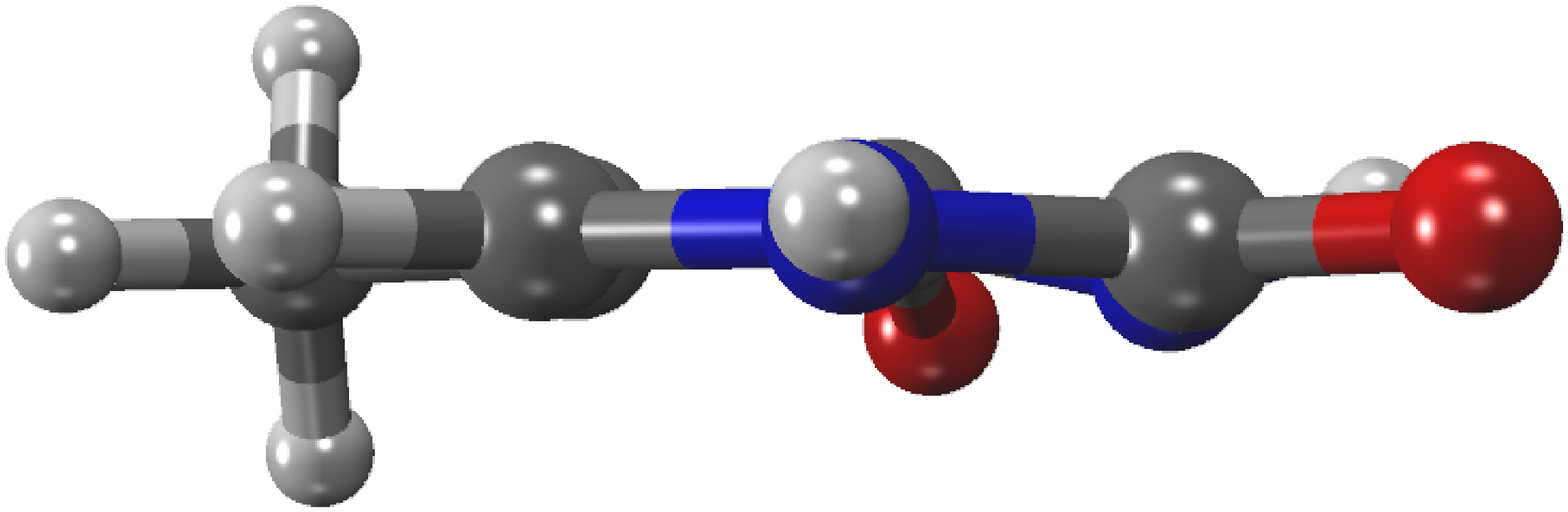}
\end{minipage}

\changeswitch

\bigskip
\noindent
\begin{minipage}{0.7\textwidth}
\small
\begin{verbatim}
 15
S1(1npi*)/S2(1pipi*) MXP (short C4-O), SA(4,3)-CASSCF(12,9)/6-31G*
C +0.106693 +0.688813 +0.038900
C -1.232698 +0.306944 +0.203065
C -0.340262 -2.032130 -0.033894
C +1.212306 -0.207715 -0.013595
C +0.405275 +2.163353 -0.007630
N -1.337394 -1.230552 +0.171881
N +0.893983 -1.590127 -0.186102
O -2.268944 +0.856316 +0.344463
O -0.529580 -3.348524 -0.103759
H -2.256217 -1.607073 +0.293248
H +1.612443 -2.266239 -0.330276
H +2.127125 +0.067920 -0.504366
H +0.804265 +2.462921 -0.980805
H -0.502267 +2.728739 +0.173192
H +1.143055 +2.446230 +0.746062
\end{verbatim}
\end{minipage}
\begin{minipage}{0.29\textwidth}
\vspace*{0.5cm}
\includegraphics[width=\textwidth]{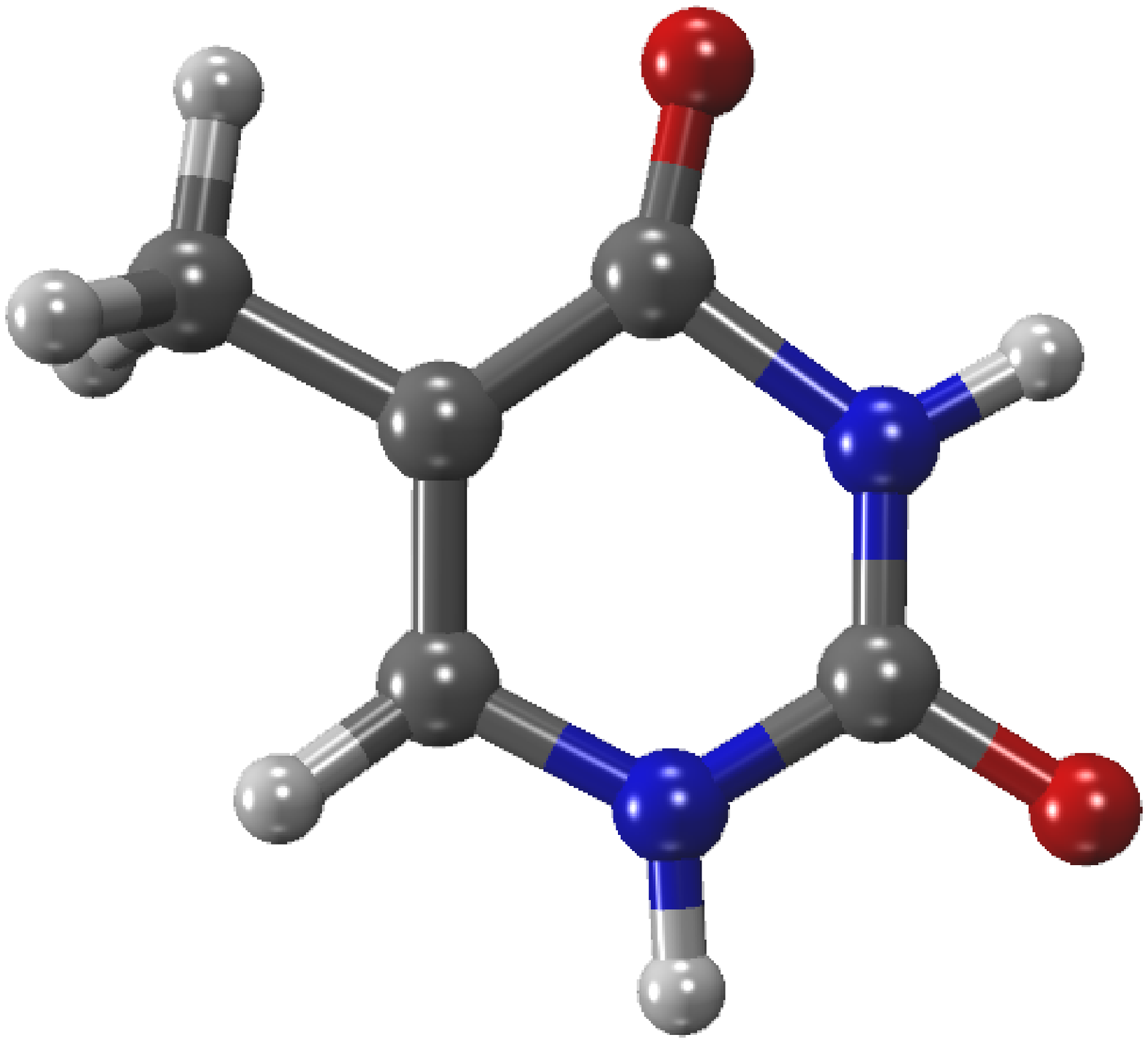}

\includegraphics[width=\textwidth]{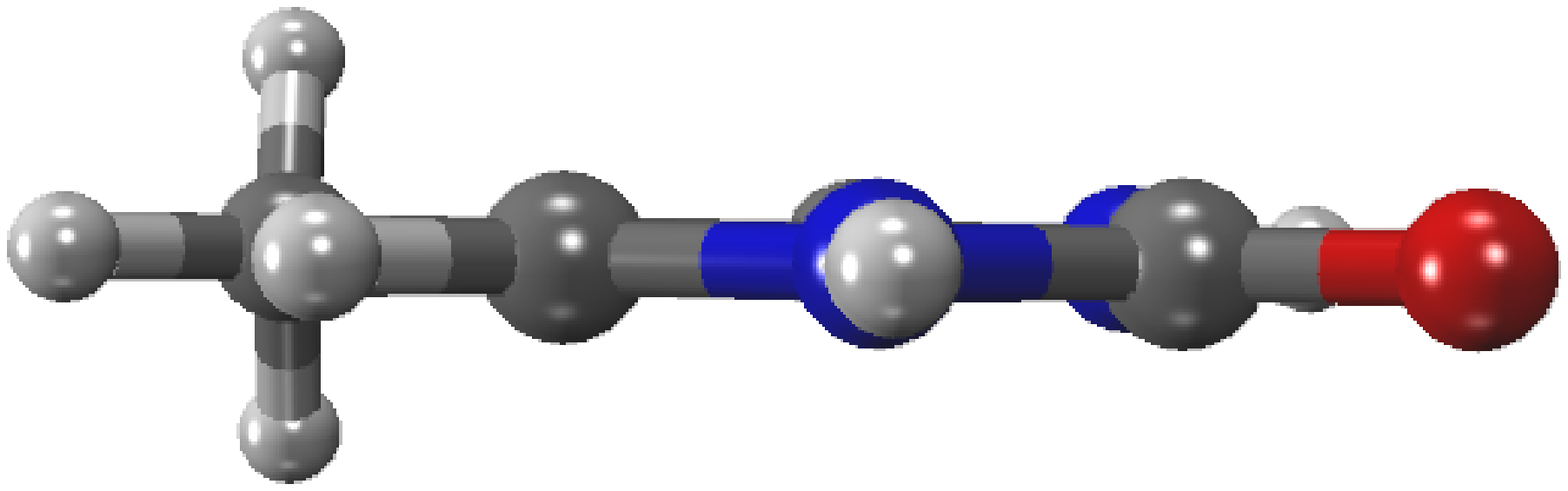}
\end{minipage}

\bigskip
\noindent
\begin{minipage}{0.7\textwidth}
\small
\begin{verbatim}
 15
S1(1npi*)/S2(1pipi*) MXP (oop-O4), SA(4,3)-CASSCF(12,9)/6-31G*
C +0.011564 +0.742077 +0.085354
C -1.309650 +0.161045 +0.048796
C -0.311824 -2.080382 +0.209688
C +1.088845 -0.162168 -0.404492
C +0.273546 +2.186433 +0.364423
N -1.390626 -1.239524 +0.076847
N +0.917867 -1.485542 +0.008556
O -1.944105 +0.821799 -0.936271
O -0.426346 -3.249633 +0.455370
H -2.268376 -1.664645 +0.277842
H +1.651361 -2.141337 -0.146732
H +1.496444 -0.016334 -1.396533
H +1.064885 +2.272661 +1.104756
H +0.606804 +2.714567 -0.525221
H -0.613520 +2.678864 +0.740597

\end{verbatim}
\end{minipage}
\begin{minipage}{0.29\textwidth}
\vspace*{0.5cm}
\includegraphics[width=0.9\textwidth]{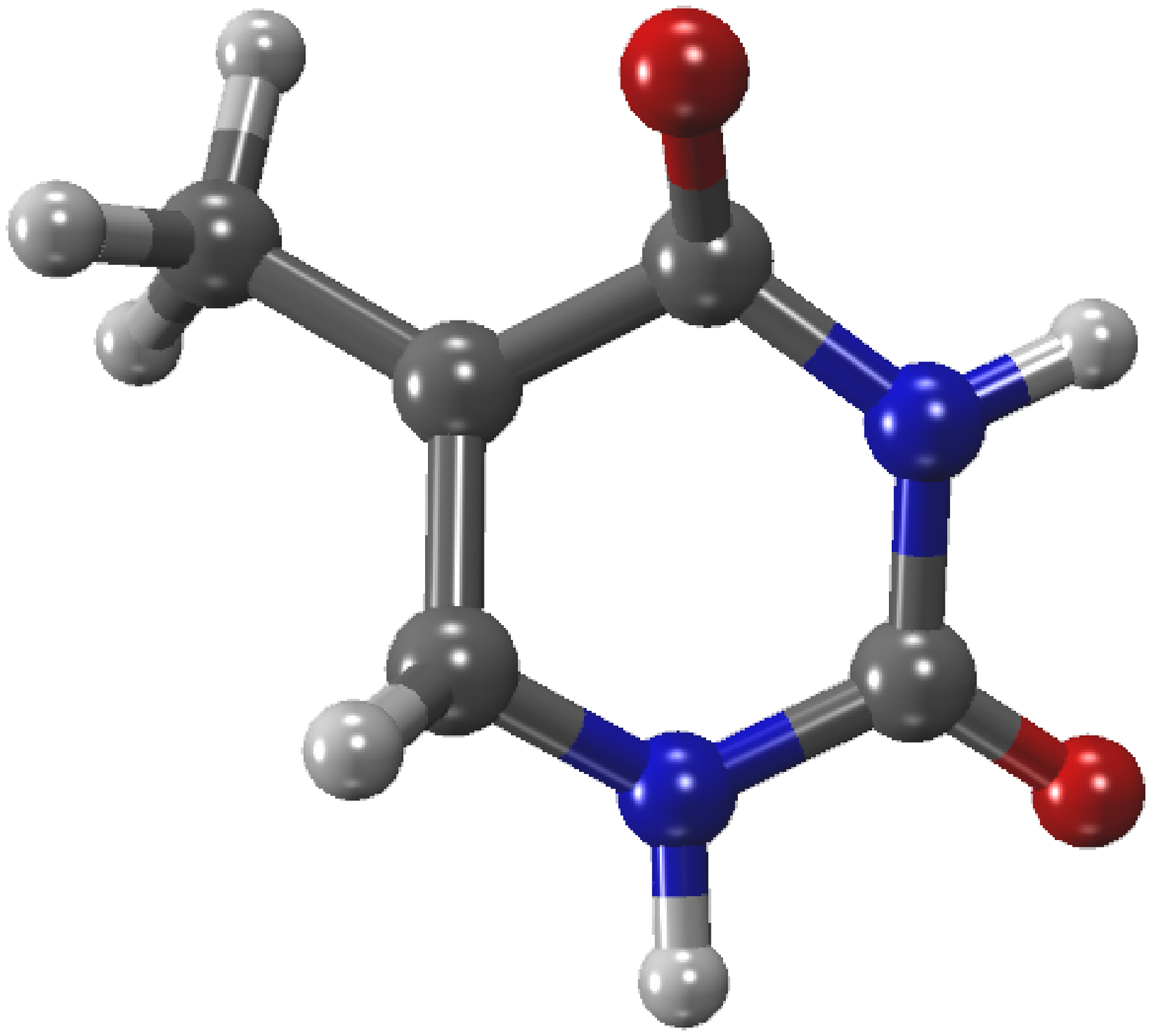}

\includegraphics[width=0.9\textwidth]{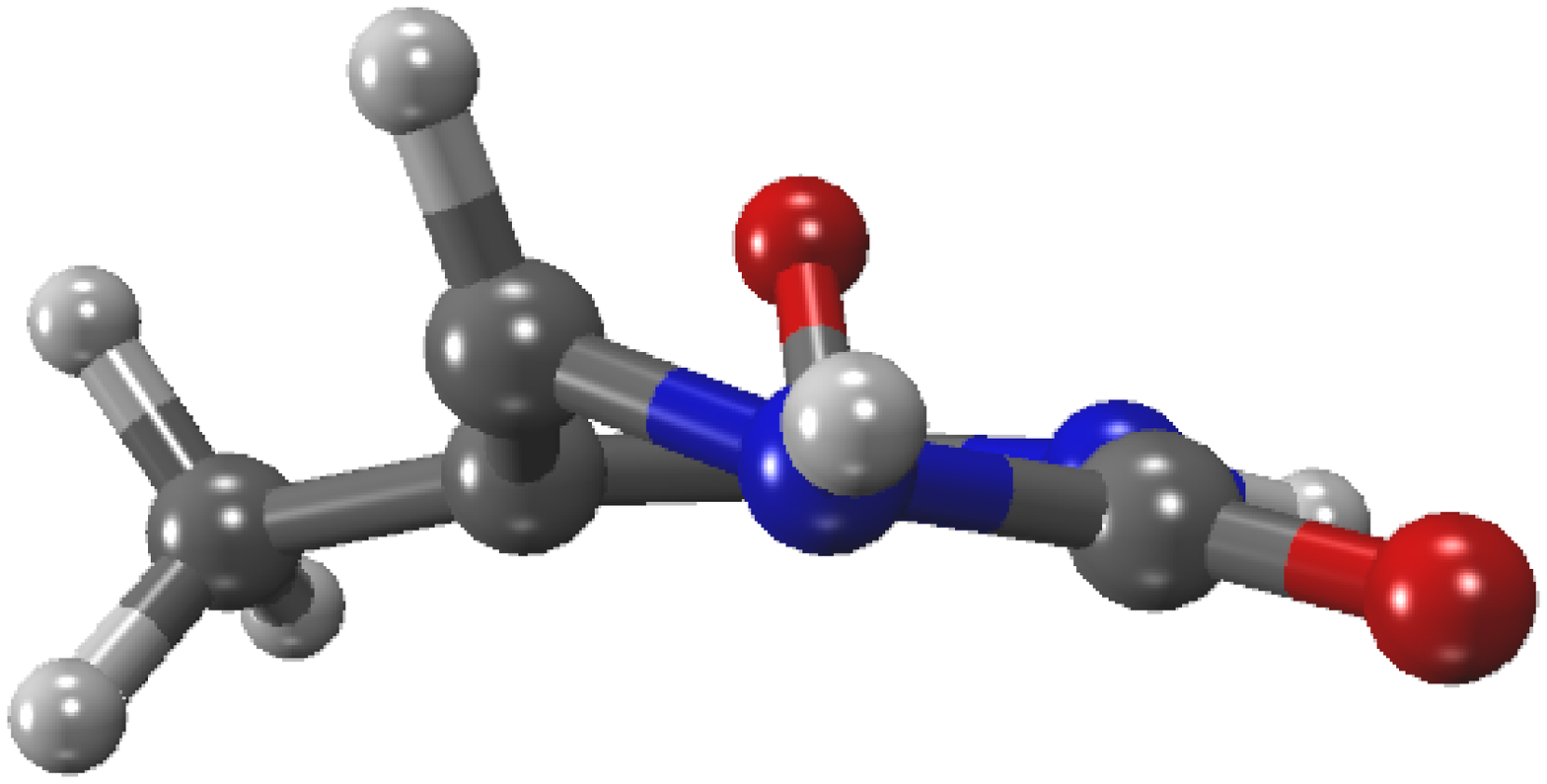}
\end{minipage}


\end{document}